\documentclass{article}
\usepackage[T1]{fontenc}
\usepackage{amsmath,amssymb}
\usepackage{geometry}
\usepackage{authblk}
\usepackage{algorithm}
\usepackage{color}
\usepackage{enumerate} 
\usepackage{amsthm} 
\usepackage[pdftex]{hyperref} 
\hypersetup{
 setpagesize=false,
 bookmarksnumbered=true,%
 bookmarksopen=true,%
 colorlinks=true,
 linkcolor=cyan, 
 citecolor=blue, 
}
\usepackage{comment}
\usepackage{braket}

\usepackage[pdftex]{graphicx,xcolor}
\usepackage{tikz}
\usepackage{amsmath}
\usetikzlibrary{arrows.meta,datavisualization}
\usepackage{pgfplots}
\pgfplotsset{compat=1.12}

\newcommand{\poly}{\mathrm{poly}}

\newtheorem{theorem}{Theorem}
\newtheorem{lemma}[theorem]{Lemma}

\theoremstyle{definition}
\newtheorem{definition}{Definition}

\newtheorem{remark2}{Remark}

\begin{document}

\title{	
		\vspace{-10truemm}
Quantum Speedup for the Minimum Steiner Tree Problem
}
\author[1]{Masayuki Miyamoto\thanks{miyamoto.masayuki.46s@st.kyoto-u.ac.jp}}
\author[2]{Masakazu Iwamura\thanks{masa@cs.osakafu-u.ac.jp}}
\author[2]{Koichi Kise\thanks{kise@cs.osakafu-u.ac.jp}}
\author[3]{Fran\c{c}ois Le Gall\thanks{legall@math.nagoya-u.ac.jp}}
\affil[1]{
Graduate School of Informatics, Kyoto University
}
\affil[2]{
Graduate School of Engineering, 
Osaka Prefecture University
}
\affil[3]{
Graduate School of Mathematics, Nagoya University
}
\date{April 9, 2019}

\maketitle

\begin{abstract}
A recent breakthrough by Ambainis, Balodis, Iraids, Kokainis, Pr\=usis and Vihrovs (SODA'19) showed how to construct faster quantum algorithms for the Traveling Salesman Problem and a few other NP-hard problems by combining in a novel way quantum search with classical dynamic programming. In this paper, we show how to apply this approach to the minimum Steiner tree problem, a well-known NP-hard problem, and
construct the first quantum algorithm that solves this problem faster than the best known classical algorithms. More precisely, the complexity of our quantum algorithm is $\mathcal{O}(1.812^k\poly(n))$, where $n$ denotes the number of vertices in the graph and $k$ denotes the number of terminals. In comparison, the best known classical algorithm has complexity  $\mathcal{O}(2^k\poly(n))$.
\end{abstract}
    
\section{Introduction}
\noindent
\paragraph{Background: Quantum speedup of dynamic programming algorithms.}
The celebrated quantum algorithm by Grover~\cite{grover1996fast} for quantum search (Grover search) gives a quadratic speed up over classical algorithms for the unstructured search problem \cite{boyer1998tight,grover1996fast}. Its generalization, quantum amplitude amplification~\cite{brassard2002quantum,mosca1998quantum}, is also useful to speed up classical algorithms. For many problems, however, Grover search or quantum amplitude amplification does not immediately give a speedup. A simple example is the Traveling Salesman Problem (TSP). The trivial brute-force algorithm for the TSP has running time $\mathcal{O}(n!)$, where~$n$ denote the number of vertices of the graph. While Grover search can be applied to improve this complexity to $\mathcal{O}(\sqrt{n!})$, the well-known classical algorithm by Held and Karp~\cite{held1962dynamic}, based on dynamic programming, already solves the TSP in $\mathcal{O}^*(2^n)$ time,\footnote{In this paper the $\mathcal{O}^*$ notation hides polynomial factors in $n$.} which is significantly better than that quantum speedup.

Recently, Ambainis, Balodis, Iraids, Kokainis, Pr\=usis and Vihrovs~\cite{ambainis2019quantum} developed a breakthrough approach to achieve quantum speedups for several fundamental NP-hard problems, by combining in a clever way Grover search and (classical) dynamic programming. For the TSP, in particular, they obtained a $\mathcal{O}^*(1.728^{n})$-time quantum algorithm, which outperforms the $\mathcal{O}^*(2^n)$-time classical algorithm mentioned above. They also constructed similar quantum algorithms, faster than the best known classical algorithms, for a few other NP-hard problems: checking the existence of a path in an hypercube (and several similar vertex ordering problems), computing the graph bandwidth, the minimum set cover problem and the feedback arc set problem. 

While the approach from \cite{ambainis2019quantum} has the potential to lead to speed-ups for other hard problems, it cannot be applied to any computational problem. The approach (currently) works only for computational problems that can be expressed with dynamic programming using a recurrence relation of a simple form. An important question is to identify which other problems can be sped-up in the quantum setting by this approach, i.e., identify which other problems admit this formulation.

In this paper we show that another fundamental NP-hard problem, the Minimum Steiner Tree Problem, can be sped up by such a combination of Grover search and dynamic programming.

\paragraph{The Minimum Steiner Tree Problem.}
Given an undirected weighted graph $G=(V,E,w)$ and a subset of terminals $K\subseteq V$, a Steiner tree is a subtree of $G$ that connects all vertices in $K$. Below, we will write $n=|V|$ and $k=|K|$. 
The task of finding a Steiner tree of minimum total weight is called the Minimum Steiner Tree problem (MST problem). This problem 
is NP-hard~\cite{karp1972reducibility}. Note that for fixed constant~$k$, this problem can be solved in polynomial time, which means that the MST problem is fixed parameter tractable~\cite{downey2012parameterized,flum2006parameterized}.

The MST problem has applications to solve problems such as power supply network, communication network and facility location problem~\cite{hwang1992steiner}. Since all these problems need to be solved in practice, designing algorithms as fast as possible for the MST problem is of fundamental importance.

A naive way to solve the MST problem is to compute all possible trees. Since the number of all trees in the graph $G=(V,E)$ can be as large as $\mathcal{O}(2^{|E|})$, this is extremely inefficient. The Dreyfus-Wagner algorithm~\cite{dreyfus1971steiner} is a well-known algorithm based on dynamic programming for solving the MST problem in time $\mathcal{O}^*(3^k)$. This algorithm has been the fastest algorithm for decades. Fuchs, Kern and Wang~\cite{fuchs2007speeding} finally improved this complexity to $\mathcal{O}^*(2.684^k)$, and M{\"o}lle, Richter and Rossmanith~\cite{molle2006faster} further improved it to $\mathcal{O}((2+\delta)^kn^{f(\delta^{-1})})$ for any constant $\delta >0$.
For a graph with a restricted weight range, Bj{\"o}rklund, Husfeldt, Kaski and Koivisto have proposed an $\mathcal{O}^*(2^k)$ algorithm using subset convolution and M{\"o}bius inversion~\cite{bjorklund2007fourier}. The main tool in all these algorithms \cite{bjorklund2007fourier,fuchs2007speeding,molle2006faster} is dynamic programming.

\paragraph{Our results.} Our main result is the following theorem (see also Table \ref{tbl:res}).
\begin{theorem}\label{th:main}
There exists a quantum algorithm that solves with high probability the Minimum Steiner Tree problem in time $\mathcal{O}^*(1.812^k)$, where $k$ denotes the size of the terminal set.
\end{theorem}
The quantum algorithm of Theorem \ref{th:main} is the first quantum algorithm that solves the MST problem faster than the best known classical algorithms. 

Our approach is conceptually similar to the approach introduced in \cite{ambainis2019quantum}: we combine Grover search and (classical) dynamic programming. All the difficulty is to find the appropriate dynamic programming formulation of the MST problem. The dynamic programming formulation used in the Dreyfus and Wagner algorithm \cite{dreyfus1971steiner} cannot be used since that characterisation of minimum Steiner trees is not suitable for Grover search. Instead, we rely on another characterization introduced by Fuchs, Kern and Wang~\cite{fuchs2007speeding}.
More precisely, Ref.~\cite{fuchs2007speeding} introduced, for any $r\ge 2$ the concept of ``$r$-split" of a graph and showed how to use it to derive a dynamic programming formulation that decomposes the computation of a minimum Steiner trees into several parts. By considering the case $r=3$, i.e., decomposing trees into three parts, they obtained their $\mathcal{O}^*(2.684^k)$-time algorithm. In Section \ref{sec:3-2} we show how to derive another recurrence relation (Equation (\ref{MST})). Interestingly, we use a 2-split to derive this relation, and not a 3-split as in \cite{fuchs2007speeding} (it seems that a 3-split only gives worse complexity in the quantum setting). We then show in Section \ref{sec:4} how to use Grover search to compute efficiently a minimum Steiner tree using Equation~(\ref{MST}). This is done by applying Grover search recursively several times with different size parameters.

\vspace{-0.3cm}
\begin{table*}[bp]
\caption{Comparison of the algorithms for the Minimum Steiner Tree problem. Here $n$ denotes the number of nodes in the graph and $k$ denotes the size of the terminal set.} \label{tbl:res}
{\centering
\begin{tabular}{| l | l |c|}
\hline Algorithm& Complexity&classical or quantum \\ \hline
Dreyfus and Wagner~\cite{dreyfus1971steiner} & $\mathcal{O}^*(3^k)$ & classical\\
Fuchs et al.~\cite{fuchs2007speeding} & $\mathcal{O}^*(2.684^k)$& classical\\
 M{\"o}lle et al.~\cite{molle2006faster} & $\mathcal{O}((2+\delta)^kn^{f(\delta^{-1})})$ & classical\\
Bj{\"o}rklund et al.~\cite{bjorklund2007fourier} 
& $\mathcal{O}^*(2^k)$\:\: [for restricted weights]& classical\\ 
\textbf{This paper} & \boldmath{$\mathcal{O}^*(1.812^k)$ } & \textbf{quantum}\\ \hline
\end{tabular}
}
\end{table*}

\section{Preliminaries}
\paragraph{General notation.}
We denote $H$ the binary entropy function, defined as 
$H(\alpha)=-\alpha\log\alpha -(1-\alpha)\log(1-\alpha)$ for any $\alpha\in[0,1]$.

\paragraph{Graph-theoretic notation.}
In this paper we consider undirected weighted graphs $G=(V,E,w)$ with weight function $w:E\rightarrow \mathbb{R}^+$, where $\mathbb{R}^+$ denotes the set of positive real numbers. Given a subset $E'\subseteq E$ of edges, we write $V(E')\subseteq V $ the set of vertices induced by $E'$, and write $w(E')=\sum_{e\in E'}w(e)$. Given a tree $T$ of $G$, i.e., a subgraph of $G$ isomorphic to a tree, we often identify $T$ with its edge set. In particular, we write its total weight $w(T)$.

\begin{figure*}[tbp]
    \centering
    \includegraphics[width=\hsize]{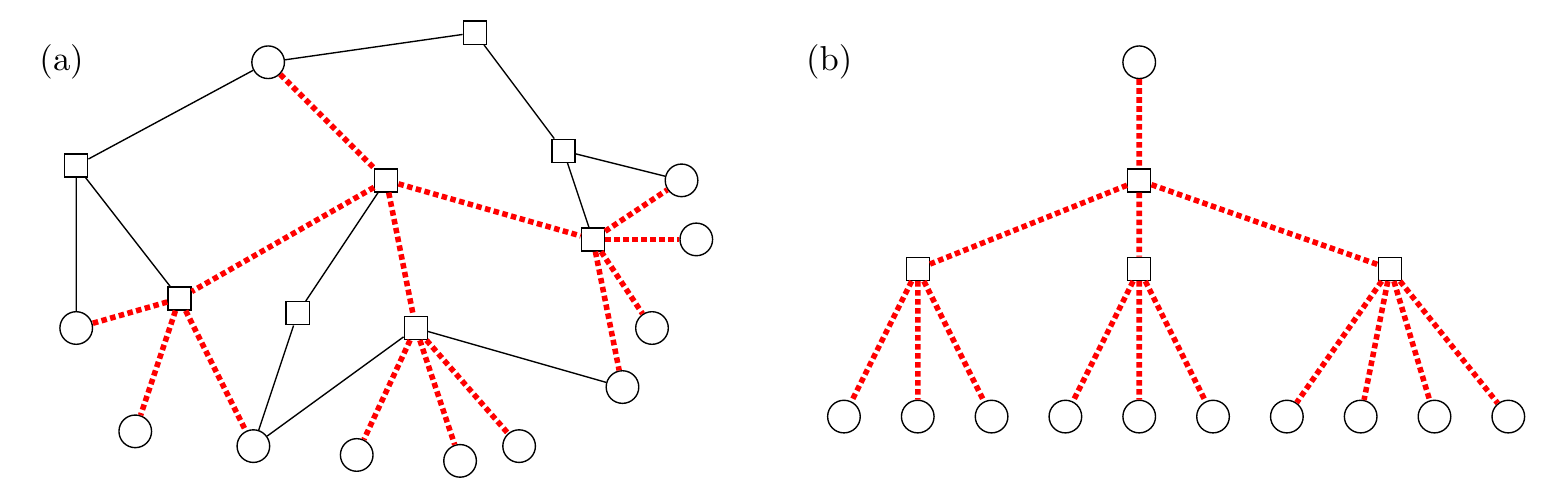}
    \caption{(a)~An example of a graph $G=(V,E,w)$. The graph is unweighted, i.e., $w(e)=1$ for all $e\in E$. Circled nodes represent the nodes in the terminal set $K$ and rectangular nodes represent the nodes in $V\setminus K$. The red dotted edges show the minimum Steiner tree $T$. In this case we have $W_G(K)=14$. (b)~The tree $T$ extracted from (a).} 
    \label{fig:example graph}
\end{figure*}

\paragraph{Minimum Steiner trees.}
Given an undirected weighted graph $G=(V,E,w)$ and a subset of vertices $K\subseteq V$, usually referred to as terminals, a Steiner tree is a tree of $G$ that spans $K$ (i.e., connects all vertices in $K$). A Steiner tree $T$ is a minimum Steiner tree (MST) if its total edge weight $w(T)$ is the minimum among all Steiner trees for $K$. Note that all leaves of a Steiner tree $T$ are necessarily vertices in $K$. We denote $W_G(K)$ the weight of an MST. Figure~\ref{fig:example graph} shows an example. The Minimum Steiner Tree Problem (MST problem) asks, given $G$ and $K$, to compute $W_G(K)$ and output an MST. In this paper we write $n=|V|$ and $k=|K|$. When describing algorithms for the MST problem, we often describe explicitly only the computation of $W_G(K)$. For all the (classical and quantum) algorithms for the MST problem described in this paper, which are all based on dynamic programming, an MST can be obtained from the computation of $W_G(K)$ simply by keeping record of the intermediate steps of the computation.

\paragraph{Graph Contraction.}
For a graph $G=(V,E,w)$ and a subset of vertices $A\subseteq V$, a graph contraction $G/A$ is a graph which is obtained by removing all edges between vertices in $A$, replacing all vertices in $A$ with one new vertex $v_A$, and replacing each edge $e\in E$ with one endpoint $u$ outside $A$ and the other endpoint in $A$ by an edge between $u$ and $v_A$ of weight $w(e)$. 
If a vertex $u\in V$ is incident to multiple edges $e_1,e_2,...,e_s\in E$ with the other endpoint in $A$, then the graph $G/ A$ has an edge $(u,v_A)$ with a weight $\min_{i=1}^sw(e_i)$ instead of having $s$ edges between $u$ and $v_A$.

\paragraph{Quantum algorithm for minimum finding.}
The quantum algorithm for mininum finding by D{\"u}rr and H{\o}yer \cite{durr1996quantum}, referred to as ``D-H algorithm" in this paper, is a quantum algorithm for finding the minimum in an (unsorted) database that is based on Grover's quantum search algorithm~\cite{grover1996fast}. More precisely, the D-H algorithm is given as input quantum access to $N$ elements $a_1,...,a_N$ from an ordered set, i.e., the algorithm has access to a quantum oracle that maps the quantum state $\ket{i}\ket{0}$ to the quantum state $\ket{i}\ket{a_i}$, for any $i\in\{1,\ldots,N\}$. The algorithm outputs with high probability (i.e., probability at least $1-1/\poly(N)$) the value $\min\{a_i|i=1,...,N\}$ using only ${\mathcal{O}}(\sqrt{N})$ calls to the oracle. This gives a quadratic speedup with respect to  classical algorithms for minimum finding.

\section{Building Blocks from Prior Work}
In this section we describe results from prior works that will be used to build our quantum algorithm. 

\subsection{The Dreyfus-Wagner algorithm}\label{sec:3-1}
The Dreyfus-Wagner algorithm \cite{dreyfus1971steiner}, referred to as ``D-W algorithm'' in this paper, solves the MST problem in time $\mathcal{O}^*(3^k)$ by using dynamic programming. 
The result from \cite{dreyfus1971steiner} that we will need in this paper is not the final algorithm, but rather the following technical result.

\begin{theorem}[\cite{dreyfus1971steiner}]\label{thm:D-W}
    For any value $\alpha\in (0,1/2]$, all the weights $W_G(X)$ for all the sets $X\subseteq K$ such that $|X|\leq \alpha |K|$ can be computed in time $\mathcal{O}^*\left( 2^{\left( H(\alpha) + \alpha \right)k} \right)$.
\end{theorem}

\begin{figure*}[tb]
    \centering
    \includegraphics[width=.7\hsize]{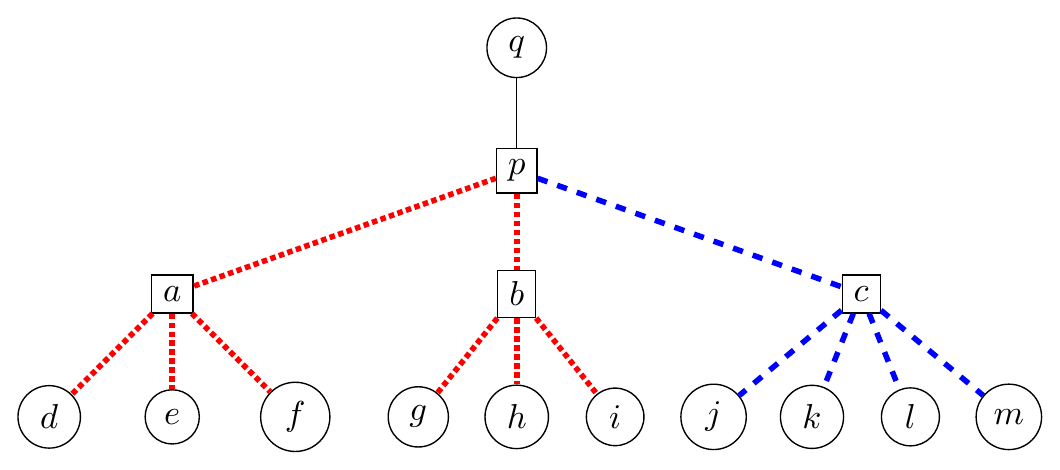}
    \caption{An illustration for Dreyfus-Wagner decomposition for the tree in Figure~\ref{fig:example graph}. The MST $T$ for $X\cup\{q\}$ is decomposed into three parts: a black solid path $P_{qp}$, an MST $T'_1$ for $X'\cup\{p\}$ with blue dashed edges, an MST $T'_2$ for $(X\backslash X')\cup\{p\}$ with red dotted edges. In this case we have $X=\{d,e,f,g,h,i,j,k,l,m\}$ and $X'=\{j,k,l,m\}$.} 
    \label{fig:DW}
\end{figure*}

For completeness we give below an overview of the proof of Theorem \ref{thm:D-W}.
The key observation is as follows.
Assume that we have an MST $T$ for $X\cup\{q\}$ where $X\subseteq K, q\in K\backslash X$. If $q$ is a leaf of $T$, then there is a vertex $p\in V(T)$ such that there is a shortest path $P_{qp}$ that connects $q$ and $p$ in $T$, and $p$ has more than two neighbors in $T$ (otherwise, $T$ is a path and we decompose $T$ into two paths).
Hence, we have $T=P_{qp}\cup T'$ where $T'$ is an MST for $X\cup\{p\}$.
Note that $p$ might not be a terminal, i.e., possibly $p\notin K$.
After removing $P_{qp}$ from $T$, $p$ splits the remaining component $T'$ into two edge disjoint subtrees, i.e., for some nontrivial subset $X'\subseteq X$, MSTs $T'_1$ for $X'\cup\{p\}$ and $T'_2$ for $(X\backslash X')\cup\{p\}$, we have the decomposition $T' = T'_1\cup T'_2$.
This holds in both cases $p\in K$ and $p\notin K$, and even when $q$ is not a leaf of $T$ (in this case, we take $p=q$ and $P_{qp}=\emptyset$). This implies that an MST $T$ for $X\cup\{q\}$ can be computed from the MSTs $T'$ for $X'\cup\{p\}$ and the shortest paths $P_{pq}$ for all $p\in V$ and $X'\subseteq X$.
%
%
We thus obtain the following recursion:
\begin{align}\label{eq:DW}
    W_G(X\cup \{q\})&=\min_{\substack{p\in V \\ X' \subset X}}\{d_G(q,p)+W_G(X'\cup \{p\}) + W_G((X\backslash X')\cup \{p\})\}
\end{align}
where $d_G(q,p)$ is the weight of a shortest path $P_{qp}$ (shortest paths of all pairs of vertices can be computed in poly$(n)$ time). See Fig.~\ref{fig:DW} for an illustration of the decomposition.

Using this recursion, weights of MSTs for all subsets of terminals $X\subseteq K$ with size $|X|\leq \alpha k$ can be computed in time
\[
    \mathcal{O}^*\left(
    \sum_{i=0}^{\alpha k} \begin{pmatrix}
      k \\ i
    \end{pmatrix} 2^i
    \right),
\]
where $\binom{k}{i}$ represents the number of sets $X\subset K$ with $|X| =i$ and $2^i$ represents the number of sets $X'\subset X$.
As claimed in Theorem~\ref{thm:D-W}, for $\alpha \leq 1/2$, this complexity is upper bounded by 
$\mathcal{O}^*\left( 2^{\left( H(\alpha) + \alpha \right)k} \right)$
    where $H$ is the binary entropy function.

\subsection{The algorithm by Fuchs, Kern and Wang}\label{sec:3-2}
Fuchs, Kern and Wang \cite{fuchs2007speeding} have improved the D-W algorithm by dividing the algorithm into two parts: a dynamic programming part and a part which merges subtrees. In this paper we will not use directly this improved algorithm. Instead, we will use the main technique introduced in \cite{fuchs2007speeding} to obtain another recurrence relation on which our quantum algorithm will be based. 

The central idea that we need is the concept of ``$r$-split'' of an MST. This concept was introduced in \cite{fuchs2007speeding} for any value $r\ge 2$ and used with $r=3$ to construct their $\mathcal{O}^*(2.684^k)$-time algorithm for the MST problem . For our purpose, on the other hand, we will need the version with $r=2$, which we define below.

\begin{definition}
Let $T$ be an MST for the terminal set $K$. A $2$-split of $T$ is an edge disjoint partition
$
    T=T_1\cup E'
$
such that $T_1$ is a subtree of $T$ and the subgraph induced by the edge subset $E'$ is a subforest of $T$.
We also use the following notation. 
\begin{align*}
    &A:=V(T_1) \cap V(E')\\
    &K_1:=K\cap V(T_1) \backslash A \\
    &K_2:=K\cap V(E') \backslash A
\end{align*}
We call $A$ the set of split nodes. When $T$ and $E'$ are both nonempty, we have $A\neq\emptyset$ since $T$ is a tree.
\end{definition}

\begin{figure*}[p]
    \centering
    \begin{tikzpicture}[scale=0.16,every node/.style={circle,draw}]

    \node (Z) at ( 3,-2) {};
    \node[black,rectangle] (A) at ( 7,-6) {};
    \node[rectangle] (B) at ( 0,-10) {};
    \node[rectangle] (C) at (8,-11) {};
    \node[rectangle] (D) at (14,-8) {};
    \draw[red,densely dotted,very thick] (7.0,-6.5) -- (8.0,-10.48);
    
    \node (E) at ( -3.5, -11) {};
    \node (F) at ( -1.5, -14.5) {};
    \node (G) at ( 2.5, -15) {};
    
    \node (H) at ( 6, -15.3) {};
    \node (I) at ( 9.5, -15.5) {};
    \node (J) at ( 11.5, -15) {};
    
    \node (K) at ( 15, -13) {};
    \node (L) at ( 16, -11) {};
    \node (M) at ( 17.5, -8) {};
    \node (N) at ( 17, -6) {};
    
    \node[rectangle] (X1) at ( 10,-1) {};
    \node[rectangle] (X2) at ( 13,-5) {};
    \node[rectangle] (X3) at (4,-10.5) {};
    \node[rectangle] (X4) at (-3.5,-5.5) {};
    
    \draw[red,densely dotted,very thick] (Z) -- (A);
    \foreach \u \v in {A/B,B/E,B/F,B/G,C/H,C/I}
        \draw[red,densely dotted,very thick] (\u) -- (\v);
    \foreach \u \v in {A/D,D/K,D/L,D/M,D/N,C/J}
        \draw[blue,dashed,very thick] (\u) -- (\v);
    \foreach \u \v in {Z/X1,D/X2,N/X2,C/G,X1/X2,X4/Z,X4/E,A/X3,X3/G,C/K,X4/B}
        \draw (\u) -- (\v);
        
    \node (Z0) at ( 41,-2) {};
    \node[rectangle] (B0) at ( 38,-10) {};
    \node[rectangle] (D0) at (52,-8) {};
    
    \node (E0) at ( 34.5, -11) {};
    \node (F0) at ( 36.5, -14.5) {};
    \node (G0) at ( 40.5, -15) {};
    
    \node (H0) at ( 44, -15.3) {};
    \node (I0) at ( 47.5, -15.5) {};
    \node (J0) at ( 49.5, -15) {};
    
    \node (K0) at ( 53, -13) {};
    \node (L0) at ( 54, -11) {};
    \node (M0) at ( 55.5, -8) {};
    \node (N0) at ( 55, -6) {};
    
    \node[rectangle] (X10) at ( 48,-1) {};
    \node[rectangle] (X20) at ( 51,-5) {};
    \node[rectangle] (X30) at (42,-10.5) {};
    \node[rectangle] (X40) at (34.5,-5.5) {};
    \node[rectangle] (PB0) at (45.7,-7) {};
    
    \draw[red,densely dotted,very thick] (Z0) -- (PB0);
    \foreach \u \v in {PB0/B0,B0/E0,B0/F0,B0/G0,PB0/H0,PB0/I0}
        \draw[red,densely dotted,very thick] (\u) -- (\v);
    \foreach \u \v in {PB0/D0,D0/K0,D0/L0,D0/M0,D0/N0,PB0/J0}
        \draw[blue,dashed,very thick] (\u) -- (\v);
    \foreach \u \v in {Z0/X10,D0/X20,N0/X20,PB0/G0,X10/X20,X40/Z0,X40/E0,PB0/X30,X30/G0,PB0/K0,X40/B0}
        \draw (\u) -- (\v);
        
    \node (Z1) at ( 3,-22) {};
    \node[black,rectangle] (A1) at ( 7,-26) {};
    \node[rectangle] (B1) at ( 0,-30) {};
    \node[rectangle] (C1) at (8,-31) {};
    
    \node (E1) at ( -3.5, -31) {};
    \node (F1) at ( -1.5, -34.5) {};
    \node (G1) at ( 2.5, -35) {};
    
    \node (H1) at ( 6, -35.3) {};
    \node (I1) at ( 9.5, -35.5) {};
    
    \draw[red,densely dotted,very thick] (Z1) -- (A1);
    \foreach \u \v in {A1/B1,A1/C1,B1/E1,B1/F1,B1/G1,C1/H1,C1/I1}
        \draw[red,densely dotted,very thick] (\u) -- (\v);

    \node[rectangle] (D2) at (52,-27) {};
    
    \node (J2) at ( 49.5, -34) {};
    \node (K2) at ( 53, -32) {};
    \node (L2) at ( 54, -30) {};
    \node (M2) at ( 55.5, -27) {};
    \node (N2) at ( 55, -25) {};
    
    \node[rectangle] (PB2) at (45.7,-27) {};
    
    \foreach \u \v in {PB2/D2,D2/K2,D2/L2,D2/M2,D2/N2,J2/PB2}
        \draw[blue,dashed,very thick] (\u) -- (\v);
    
    \node (Z3) at ( 24,-48) {};
    \node[black,rectangle] (A3) at ( 28,-52) {};
    \node[rectangle] (B3) at ( 21,-56) {};
    \node[rectangle] (C3) at (29,-57) {};
    \node[rectangle] (D3) at (35,-54) {};
    
    \node (E3) at ( 17.5, -57) {};
    \node (F3) at ( 20.5, -59.5) {};
    \node (G3) at ( 23.5, -61) {};
    
    \node (H3) at ( 27, -61.3) {};
    \node (I3) at ( 30.5, -61.5) {};
    \node (J3) at ( 32.5, -61) {};
    
    \node (K3) at ( 36, -59) {};
    \node (L3) at ( 37, -57) {};
    \node (M3) at ( 38.5, -54) {};
    \node (N3) at ( 38, -52) {};

    \draw[red,densely dotted,very thick] (Z3) -- (A3);
    \foreach \u \v in {A3/B3,A3/C3,B3/E3,B3/F3,B3/G3,C3/H3,C3/I3}
        \draw[red,densely dotted,very thick] (\u) -- (\v);
    \foreach \u \v in {A3/D3,D3/K3,D3/L3,D3/M3,D3/N3,C3/J3}
        \draw[blue,dashed,very thick] (\u) -- (\v);
        
    \draw [->] (6.5,-18) -- (6.5,-22.5);
    \draw [->] (50,-18) -- (50,-22.5);
    
    \draw [->] (50.5,-38) -- (28.5,-40);
    \draw [->] (5.5,-38) -- (25.5,-40);
    \node[draw=none,fill=none] at ( 27, -40) {$\bigoplus$};
    \draw[->] (27,-42) -- (27,-46);
    
    \draw[double,<->] (6.5,7) -- (6.5,0);
    \draw[double,<->] (45,7) -- (45,0);

    \node (Z4) at ( 7,25) {$a$};
    \node[black,rectangle] (A4) at ( 7,20) {$b$};
    \node[rectangle] (B4) at ( -4,15) {$c$};
    \node[rectangle] (C4) at (7,15) {$d$};
    \node[rectangle] (D4) at (18,15) {$e$};
    
    \node (E4) at ( -7.5, 10) {$f$};
    \node (F4) at ( -4, 10) {$g$};
    \node (G4) at ( -0.5, 10) {$h$};
    
    \node (H4) at ( 3, 10) {$i$};
    \node (I4) at ( 6.5, 10) {$j$};
    \node (J4) at ( 10, 10) {$k$};
    
    \node (K4) at ( 13.5, 10) {$l$};
    \node (L4) at ( 17, 10) {$m$};
    \node (M4) at ( 20.5, 10) {$n$};
    \node (N4) at ( 24, 10) {$o$};
    \draw[red,densely dotted,very thick] (Z4) -- (A4);
    \foreach \u \v in {A4/B4,A4/C4,B4/E4,B4/F4,B4/G4,C4/H4,C4/I4}
        \draw[red,densely dotted,very thick] (\u) -- (\v);
    \foreach \u \v in {A4/D4,D4/K4,D4/L4,D4/M4,D4/N4,C4/J4}
        \draw[blue,dashed,very thick] (\u) -- (\v);
 
    \node (Z5) at ( 45.5,25) {$a$};
    \node[rectangle] (B5) at ( 34.5,18) {$c$};
    \node[rectangle] (C5) at (45.5,18) {$v_{\{b,d\}}$};
    \node[rectangle] (D5) at (56.5,18) {$e$};
    
    \node (E5) at ( 31, 10) {$f$};
    \node (F5) at ( 34.5, 10) {$g$};
    \node (G5) at ( 38, 10) {$h$};
    \node (H5) at ( 41.5, 10) {$i$};
    \node (I5) at ( 45, 10) {$j$};
    \node (J5) at ( 48.5, 10) {$k$};
    
    \node (K5) at ( 52, 10) {$l$};
    \node (L5) at ( 55.5, 10) {$m$};
    \node (M5) at ( 59, 10) {$n$};
    \node (N5) at ( 62.5, 10) {$o$};
        \draw[red,densely dotted,very thick] (Z5) -- (C5);
    \foreach \u \v in {C5/B5,B5/E5,B5/F5,B5/G5,C5/H5,C5/I5}
        \draw[red,densely dotted,very thick] (\u) -- (\v);
    \foreach \u \v in {C5/D5,D5/K5,D5/L5,D5/M5,D5/N5,C5/J5}
        \draw[blue,dashed,very thick] (\u) -- (\v);
        
    \node[draw=none,fill=none] at (-10,25)  {(a)};
    \node[draw=none,fill=none] at (33,25)  {(b)};    
    \node[draw=none,fill=none] at (-10,0)   {(c)};
    \node[draw=none,fill=none] at (33,0)   {(d)};
    \node[draw=none,fill=none] at (-10,-21) {(e)};
    \node[draw=none,fill=none] at (33,-21) {(f)};
    \node[draw=none,fill=none] at (13,-47)  {(g)};
    \draw[dashed] (27,30) -- (27,-37);
    \node[rectangle] at (47.5,31) {the graph $G/\{b,d\}$};
    \node[rectangle] at (5,31) {the graph $G$};
\end{tikzpicture}
    \caption{An example of $2$-split $T= T_1\cup E'$. In this graph we have the terminal set $K=\{a,f,g,h,i,j,k,l,m,n,o\}$, $K_1 = \{ f,g,h,i,j\}$, and $A=\{b,d\}$.  (a):~The red dotted edges show the tree $T_1$ and the blue dashed edges show the forest $E'$. (b):~The contracted graph $G/\{b,d\}$. (c):~Graph $G$ containing the tree of (a). (d):~Graph $G/\{b,d\}$ containing the tree of (b). (e):~The tree induced by red dotted edges in graph $G$. (f):~The tree induced by blue dashed edges in graph $G/\{b,d\}$. (g):~The minimum Steiner tree. This is obtained by merging the tree with red dotted edges of (e) extracted from $G$ and the tree with blue dashed edges of (f) extracted from $G/\{b,d\}$.} 
    \label{fig:Fuchs flowchart}
\end{figure*}
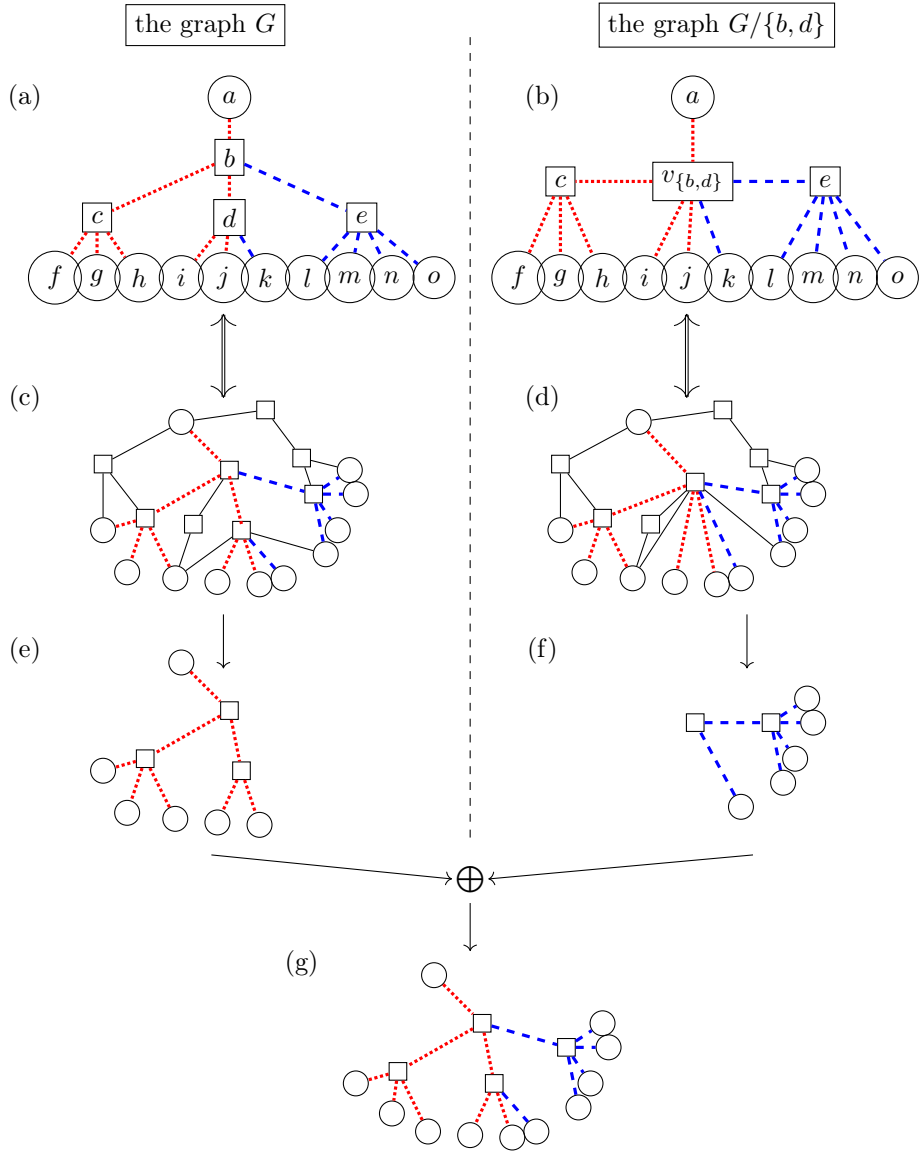

We use the following two results from~\cite{fuchs2007speeding} (see also Fig.~\ref{fig:Fuchs flowchart} for an illustration).

\begin{lemma}[\cite{fuchs2007speeding}]\label{theo:1}
Let $T$ be an MST for the terminal set $K$.
    For any 2-split $T=T_1\cup E'$, the following two properties hold:
    \begin{enumerate}
      \item[$\bullet$]  In the graph $G$, the tree $T_1$ is an MST for $K_1 \cup A$;
      \item[$\bullet$]  In the graph $G/A$, the subgraph $E'/A$ (i.e., the result of contracting $A$ in the subgraph of $G$ induced by $E'$) is an MST for $K_2\cup \{v_{A} \}$ where $v_A$ denotes the added vertex introduced in $G/A$ during the contraction.
    \end{enumerate}

\end{lemma}
\begin{theorem}[\cite{fuchs2007speeding}]\label{theo:2}
Let $T$ be an MST for the terminal set $K$.
    For any $\eta >0$ and any $0<\alpha\leq \frac{1}{2}$, there exists a 2-split $T=T_1\cup E'$ such that the following two conditions hold:
    \begin{itemize}
    \item[$\bullet$] 
    $(\alpha - \eta )k\leq |K_1| \leq (\alpha + \eta )k$; 
    \item[$\bullet$] 
    and $|A|\le
    \lceil \log{(1/\eta)} \rceil$.
    \end{itemize}
\end{theorem}

By Lemma \ref{theo:1} and Theorem \ref{theo:2}, we obtain the following recursion for any $\eta >0$ and any $0<\alpha\leq \frac{1}{2}$:
\begin{align}\label{MST}
    W_G(K)=\min_{\substack{K_1\subseteq K\\ |K_1|= (\alpha\pm\eta)k}} \  \min_{\substack{A\subseteq V \\|A|\leq \lceil\log(1/\eta)\rceil}}
    \Big\{ W_G(K_1\cup A)+W_{G/A}(K_2\cup \{v_{A} \}) \Big\},
\end{align}
where $K_2$ is defined from $K$ and $A$ as $K_2=K \backslash (K_1\cup A)$. In Equation (\ref{MST}) the shorthand ``$|K_1|=(\alpha \pm \eta)k$'' means $(\alpha - \eta)k \leq |K_1| \leq (\alpha + \eta)k$ and $W_{G/A}(K_2 \cup \{v_A\})$ is the weight of an MST for $K_2 \cup \{v_A\}$ in the graph $G/A$. 

%
%
%
\section{Quantum Algorithm for the MST}\label{sec:4}
In this section we present our quantum algorithm for the MST. The main idea is to recursively apply the D-H algorithm on Equation (\ref{MST}).

\subsection{Our Quantum Algorithm}

\begin{figure}[tbp]\label{Alg1}
\begin{algorithm}[H]
input: a graph $G=(V,E,w)$ and a subset of vertices $K\subseteq V$\\
parameters: two constants $\beta\in(0,1/2]$ and $\varepsilon\in (0,1)$\\
output: a minimum Steiner tree for $K$ in $G$.
\begin{enumerate}
\item 
For all $ X\subseteq K$ such that $|X| \leq ((1-\beta)/4 + 15 \varepsilon )k$  and all $A\subseteq V$ such that $|A| \leq \lceil \log (1/\varepsilon)\rceil$, compute the values of $W_G(X\cup A)$ and $W_{G/A}(X\cup\{v_A\})$ classically using the D-W algorithm. 
\item 
    Apply the D-H algorithm to Eq.~(\ref{MST}) three times recursively. In the last recursive call, directly use the values computed at Step 1.
\end{enumerate}
\caption{Quantum algorithm for \textsc{Minimum Steiner Tree}}
\end{algorithm}
\end{figure}

Algorithm~1 shows our quantum algorithm, which consists of a classical part (Step 1) and a quantum part (Step 2). It uses two parameters $\beta\in(0,1/2]$ and $\varepsilon\in (0,1)$. The value of $\beta$ will be set in the analysis of Section \ref{subsec:4.1}, and $\varepsilon$ will be a very small constant.

At Step 1, the algorithm computes the values of $W_G(X\cup A)$ and $W_{G/A}(X\cup\{v_A\})$ for all $ X\subseteq K$ such that $|X| \leq ((1-\beta)/4 + 15 \varepsilon )k$  and all $A\subseteq V$ such that $|A| \leq \lceil \log (1/\varepsilon)\rceil$. (Remember the definition of $v_A$ in Lemma \ref{theo:1}). 
This is done classically, using the D-W algorithm.

At Step 2, we use D-H algorithm on Equation (\ref{MST}), three times recursively, to compute a minimum Steiner tree for $K$. 
Let us now describe more precisely how Step 2 is implemented. 
The three levels of application of the D-H algorithm in our algorithm use Equation (\ref{MST}) in a slightly different way:
\begin{itemize}
    \item Level 1: D-H algorithm over Equation (\ref{MST}) with parameters $\alpha = 1/2$ and $\eta=\varepsilon$. This implements a search over all $K_1\subset K$ such that $|K_1|=(\frac{1}{2}\pm \varepsilon)k$ and all $A\subseteq V$ such that $|A|\leq \lceil\log(1/\varepsilon)\rceil$. This requires procedures computing $W_G(K_1\cup A)$ and  $W_{G/A}(K_2\cup \{v_{A} \})$, where $K_2=K\backslash(K_1\cup A)$. These two procedures are implemented at Level 2.
    \item Level 2: D-H algorithm over each of the following two formulas, which are obtained using Equation (\ref{MST}) with parameters $\alpha = 1/2$ and $\eta=\varepsilon$.
    \begin{align*}
            &\hspace{-4mm}W_G(K_1\cup A)=\\ 
            &\hspace{6mm}\min_{\substack{K_2\subseteq K_1\cup A\\ |K_2|= (\frac{1}{4}\pm O(\varepsilon))k}} \  \min_{\substack{A'\subseteq V \\|A'|\leq \lceil\log(1/\varepsilon)\rceil}}
            \Big\{ W_G(K_2\cup A')+W_{G/A'}(K_3\cup \{v_{A'} \}) \Big\},
    \end{align*} where $K_3= (K_1\cup A)\backslash (K_2\cup A')$.
    \begin{align*}
            &W_{G/A}(K_2\cup \{v_A\})=\\
            &\hspace{6mm}\min_{\substack{K_4\subseteq K_2\cup \{v_A\}\\ |K_4|= (\frac{1}{4}\pm O(\varepsilon))k}} \  \min_{\substack{A'\subseteq V \\|A'|\leq \lceil\log(1/\varepsilon)\rceil}}
            \Big\{ W_{G/A}(K_4\cup A')+W_{(G/A)/A'}(K_5\cup \{v_{A'} \}) \Big\},
    \end{align*} where $K_5= (K_2\cup \{v_A\})\backslash (K_4\cup A')$.
This requires procedures computing the four quantities
\begin{align*}W_G(K_2\cup A'),\: W_{G/A'}(K_3\cup \{v_{A'} \}),\: W_{G/A}(K_4\cup A'),\: W_{(G/A)/A'}(K_5\cup \{v_{A'} \}). \end{align*}
These four procedures are implemented at Level 3.    
    \item Level 3: D-H algorithm over each of the four corresponding formulas, which are obtained from Equation (\ref{MST}), with parameter $\alpha=\beta$, $\alpha=(1-\beta)$, $\alpha=\beta$ and $\alpha=(1-\beta)$, respectively, for some $\beta\in(0,1/2]$, and parameter $\eta=\varepsilon$.
    For example, the first formula, which corresponds to the computation of the term $W_G(K_2\cup A')$, is:
    \begin{align*}
        &W_G(K_2\cup A')=\\
        &\hspace{5mm}\min_{\substack{K_6\subseteq K_2\cup A'\\ |K_6|= (\frac{\beta}{4}\pm O(\varepsilon))k}} \  \min_{\substack{A''\subseteq V \\|A''|\leq \lceil\log(1/\varepsilon)\rceil}}
            \Big\{ W_G(K_6\cup A'')+W_{G/A''}(K_7\cup \{v_{A''} \}) \Big\},
    \end{align*} 
    where $K_7=(K_2\cup A')\backslash (K_6\cup A'')$.
    This time, the quantities $W_G(K_6\cup A'')$ and $W_{G/A''}(K_7\cup \{v_{A''} \})$ in this formula (and similarly for the other three formulas) can be obtained directly from the values computed at Step 1 of the algorithm.\footnote{Indeed, it is easy to check that all the $O(\varepsilon)$ terms in the above analysis are actually upper bounded by $15\varepsilon$.}
\end{itemize}

\subsection{Running Time}\label{subsec:4.1}
The parameter $\varepsilon$ is a small constant. To simplify the analysis below we introduce the following notation: the symbol $\hat{\mathcal{O}}$ hides all factors that are polynomial in $n$ and also all factors of the form $2^{O(\varepsilon k)}$.  
\paragraph{Analysis of the classical part.}
Note that constructing the contracted graphs $G/A$ can be done in polynomial time.
By using Theorem~\ref{thm:D-W},  the complexity of the classical part of the algorithm is
\begin{align}\label{eq:classical}
    \hat{\mathcal{O}}
    \left( 
    \begin{pmatrix}
         k  \\
         (1-\beta)k/4 
    \end{pmatrix}
    2^{(1-\beta)k/4 } 
    \right) = \hat{\mathcal{O}}\left( 2^{\left( H(\frac{1-\beta}{4}) +\frac{1-\beta}{4}  \right)k} \right).
\end{align}

\paragraph{The quantum part.}
At step 2 of our algorithm, we apply the D-H algorithm in three levels. The size of the search space of the D-H algorithm executed at Level~1 is
\begin{align}\label{eq:size1}
   \hat{\mathcal{O}}\left( \begin{pmatrix}
         k   \\
         k/2 
    \end{pmatrix} \right).
\end{align}
The size of the search space of each of the two executions of the D-H algorithm at Level~2 is 
\begin{align}\label{eq:size2}
   \hat{\mathcal{O}}\left( \begin{pmatrix}
         k/2   \\
         k/4 
    \end{pmatrix} \right).
\end{align}
The size of the search space of each of the four executions of the D-H algorithm at Level~3 is
\begin{align}\label{eq:size3}
   \hat{\mathcal{O}}\left( \begin{pmatrix}
         k/4   \\
         \beta k/4 
    \end{pmatrix} \right),
\end{align}
respectively.
The complexity of the quantum part of this algorithm is thus
\begin{align}\label{eq:quantum}
    \hat{\mathcal{O}}\left( 
    \sqrt{
    \begin{pmatrix}
         k  \\
         k/2 
    \end{pmatrix}
    \begin{pmatrix}
         k/2   \\
         k/4 
    \end{pmatrix}
    \begin{pmatrix}
         k/4   \\
         \beta k/4 
    \end{pmatrix}
    }
    \right).
\end{align}
\paragraph{Analysis of the parameter $\beta$.}
Using Stirling's Formula, the classical and quantum parts of the complexity (Equations (\ref{eq:classical}) and (\ref{eq:quantum})) can be respectively expressed as
\begin{align}
 \hat{\mathcal{O}}\left( 2^{\left( H(\frac{1-\beta}{4}) +\frac{1-\beta}{4}  \right)k} \right)
\text{ and }
\hat{\mathcal{O}}\left(2^{\frac{1}{2}\left( \frac{3}{2}+\frac{H(\beta)}{4} \right)k}\right).
\end{align}
Since the complexity is minimized when the complexities of the classical and quantum parts equal, we can optimize the parameter $\beta$ by solving the following equation:
\begin{align}\label{eq:parameter}
   H\left(\frac{1-\beta}{4}\right) +\frac{1-\beta}{4} =
   \frac{1}{2}\left( \frac{3}{2}+ \frac{H(\beta)}{4} \right).
\end{align}

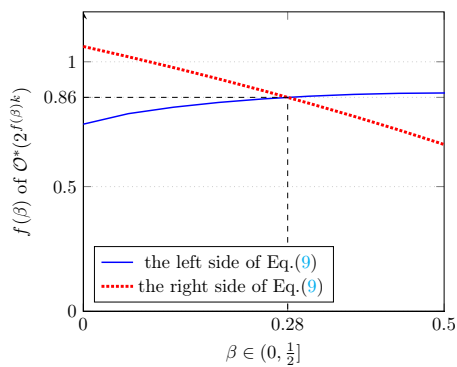
\begin{figure}[tbp]
  \begin{center}
   \begin{tikzpicture}[domain=0:1.5,scale=0.7]
 \begin{axis}[
    xlabel={$\beta\in(0,\frac{1}{2}]$},
    ylabel={$f(\beta)$ of $\mathcal{O}^*(2^{f(\beta)k})$},
    xmin=0, xmax=0.5,
    ymin=0, ymax=1.2,
    xtick={0,0.28325,0.5},
    ytick={0,0.5,0.8574,1.0},
    legend pos=south west, 
    ymajorgrids=true,
    grid style=dotted,
]
   \draw [thick, -stealth](-0.5,0)--(0.7,0) node [anchor=north]{$\beta$};
   \draw [thick, -stealth](0,-0.1)--(0,1.2) node [anchor=east]{$y$};

   
   \addplot[color=blue, thick]{{(3/2 + \x*ln(1/\x)/(4*ln(2)) + (1-\x)*ln(1/(1-\x))/(4*ln(2)) )/2 }};
    \addlegendentry{the left side of Eq.(\ref{eq:parameter})}
    
   \addplot[color=red,ultra thick, densely dotted]{{(1-\x)*ln(4/(1-\x))/(4*ln(2)) + (3+\x)*ln(4/(3+\x))/(4*ln(2)) + (1-\x)/4}};
    \addlegendentry{the right side of Eq.(\ref{eq:parameter})}

   \draw [dashed](0,0.8574) node [anchor=east]{$0.8574$}--(0.28325,0.8574)--(0.28325,0) node[anchor=north]{$0.28325$};
   
   \end{axis}
 
\end{tikzpicture}
  \end{center}
  \caption{Running time of our algorithm.} 
  \label{fig:graph}
\end{figure}
Numerical calculation show that the solution of this equation is $\beta \approx 0.28325$, which gives total running time $\hat{\mathcal{O}}(c^k)$ for $c=1.8118...$ (see also Figure~\ref{fig:graph}).
By taking an appropriately small choice of $\varepsilon$, we thus obtain running time $\mathcal{O}^*(1.812^k)$, as claimed in Theorem \ref{th:main}. 

\begin{remark2}
As we discuss in Appendix \ref{subsec:4.2}, introducing additional parameters in level 1 or level 2 of Step 2 (instead of using $\alpha=1/2$) does not improve the running time. Modifying the number of levels (e.g., using two levels, or four levels) also leads to a worse complexity.
\end{remark2}

\section*{Acknowledgements}
The authors are grateful to Shin-ichi Minato for his support. FLG was supported by JSPS KAKENHI grants Nos.~JP16H01705, JP19H04066, JP20H00579, JP20H04139 and by the MEXT Quantum Leap Flagship Program (MEXT Q-LEAP) grant No.~JPMXS0118067394. 

\bibliographystyle{plain} 
\bibliography{main}

\appendix
\section{Optimizing the number of levels of the D-H algorithm \label{subsec:4.2}}
\noindent
In Algorithm~1, we use the D-H algorithm recursively in three levels (Level 1, Level 2, and Level 3). In this appendix we show that this approach is essentially optimal, i.e., we show that using $\ell\le 2$ levels or $\ell\ge 4$ levels only leads to worse complexity. 
More precisely, we show that applying the D-H algorithm in $\ell\ge 4$ levels increases the complexity of the quantum part, while applying the D-H algorithm in $\ell\le 2$ levels increases the complexity of the classical part. 

Consider the case of $\ell$ levels, for $\ell\ge 1$. It is easy to see that the optimal choice for the parameter $\alpha$ in Equation (\ref{MST}) is always $\alpha=1/2$ except possibly at the last level. We denote $\beta\in(0,1/2]$ the parameter used at the last level. The complexity of the classical part of the algorithm is
\begin{align}\label{level classical}
    &\hat{\mathcal{O}}\left(
        \left( \begin{array}{c}
         k  \\
         (1-\beta) k/2^{\ell-1}
    \end{array} \right)2^{\frac{1-\beta}{2^{\ell-1}}k}\right)
    =\hat{\mathcal{O}}\left( 2^{\left( H\left(\frac{1- \beta}{2^{\ell-1}} \right) + \frac{1-\beta}{2^{\ell-1}} \right)k} \right).
\end{align}
The complexity of the quantum part of the algorithm is
\begin{align}\label{level quantum}
    \hat{\mathcal{O}}\left( \sqrt{
    \left(\begin{array}{c}
         k  \\
         k/2
    \end{array}\right)\left(\begin{array}{c}
         k/2  \\
         k/2^2
    \end{array}\right)
    \cdots
    \left(\begin{array}{c}
         k/2^{\ell-2}  \\
         k/2^{\ell-1}
    \end{array}\right)
    \left(\begin{array}{c}
         k/2^{\ell-1}  \\
         \beta k/2^{\ell-1}
    \end{array}\right)
    } \right) 
    =\hat{\mathcal{O}}\left( 2^{(1-\frac{1}{2^{\ell-1}} + \frac{H\left(\beta\right)}{2^{\ell-1}})k}  \right).
\end{align}
Table~\ref{tbl:levels} shows the complexity of the classical and quantum parts as a function of~$\ell$. 
For the case of $\ell\le 2$, the complexity of the classical part exceeds $\hat{\mathcal{O}}(2^k)$. The complexity of the quantum part is at least 
\[
\hat{\mathcal{O}}\left(2^{1-\frac{1}{2^{\ell-1}}k}\right).\] 
Even when $\ell=4$, this complexity is $\hat{\mathcal{O}}\big(2^{\frac{7}{8}k}\big) = \mathcal{O}^*\left( 1.835^k \right)$, which is worse than the complexity we obtain for $\ell=3$.

\begin{table*}[t]
\caption{Comparison of the running time of algorithms that use D-H algorithm in $\ell$ levels.} \label{tbl:levels}
\begin{tabular}{| c | c |c|c|c|c|}
\hline $\ell$ &1&2&3&$\geq$ 4\\ \hline
Complexity given by Eq.~(\ref{level classical})& $\:\hat{\mathcal{O}}(2^{1.5k})\:$ &$\:\hat{\mathcal{O}}(2^{1.062k})\:$ & $\:\hat{\mathcal{O}}(1.812^k)\:$& --- \\
Complexity given by Eq.~(\ref{level quantum})& $\:\hat{\mathcal{O}}(2^{0.5k})\:$ & $\:\hat{\mathcal{O}}(2^k)\:$ &$\:\hat{\mathcal{O}}(1.812^k)\:$ & at least $\hat{\mathcal{O}}(2^{1-\frac{1}{2^{l-1}}k})$ \\ \hline
     optimal values of $\beta$ & 1/2 & $1/2$ &  $\approx$ 0.28325 & --- \\ \hline
\end{tabular}
\end{table*}

\end{document}